\documentclass[conference]{IEEEconf}

\input epsf
\usepackage{graphicx}
\usepackage{multirow}

\usepackage{makecell}
\usepackage{floatrow}
\usepackage{url}

\usepackage[ruled,linesnumbered]{algorithm2e}
\begin{document}

\title{\textbf{\Large PTSG: a test generation tool based on extended finite state machine\\}}

\author{Zhijie Pan, Ting Shu$^{*}$, Zuohua Ding\\
	\normalsize School of Computer Science and Technology, Zhejiang Sci-Tech University, Hangzhou 310018, China\\
	\normalsize panpan\_zj@163.com, shuting@zstu.edu.cn, zuohuading@zstu.edu.cn\\
	\normalsize *shuting@zstu.edu.cn
}


\maketitle
\begin{abstract}
The Extended Finite State Machine (EFSM) is one of the most popular modeling approaches for model-based testing. However, EFSM-based test case generation is susceptible to the infeasible (inexecutable) path problem, which stems from the conflict of predicates (guards) between transitions in the path. Therefore, in order to derive feasible test cases, a test generation algorithm needs to dynamically acquire information about the model and verify the feasibility of the generated test path through the simulation execution of the model. The traditional method of constructing executable models using hard-coding for different EFSM models under test has limitations such as inflexibility, time-consuming and error-prone. To address this issue, this paper develops an open source test generation tool for testing EFSM-specified systems, PTSG, to support the automatic generation of executable test cases. It decouples the EFSM model description, parsing and simulation execution functions from the test generation algorithm, which can effectively improve the efficiency and quality of test generation. In particular, PTSG first uses a well-designed JSON syntax to describe the specific EFSM under test. Then, based on the model description file, it uses lexical and syntactic parsers to dynamically extract model information to construct various model objects in memory such as state configurations, transitions, etc. Finally, the tool provide a series of service interfaces to support model information acquisition, transition feasibility evaluation, and model simulation execution. A case study of test sequence generation for the SCP protocol model demonstrates the capability and promise of the PTSG to serve executable test cases.
\end{abstract}

%
\IEEEpeerreviewmaketitle

\section{Introduction}
Software plays an increasingly important role in today's society. The correctness of software is a prerequisite for its applicability. However, potential defects in software can lead to serious failures, which can have catastrophic consequences such as causing economic losses, endangering human life and the environment, etc. For example, bugs in the autopilot software caused the fatal Tesla crash \cite{banks2018driver}. In order to minimize bugs and improve software quality, software testing has become an indispensable part of the software development life cycle. Model-based testing (MBT) \cite{dias2007survey} is a typical testing method, where test cases are automatically derived from a system’s model. The model describes the expected behavior of the software under test and can be employed to support automated test case generation. Among the available test models, Extended Finite State Machine (EFSM) is widely used \cite{kalaji2011integrated}\cite{yang2015efsm}\cite{dssouli2017testing}, owing to its simplicity and ease of understanding, and the ability to portray both control and data flows of the system.

In practice, EFSM-based testing approaches need to have the functions such as model description, dynamic simulation of model behavior, and model information extraction. In addition, the infeasible test path problem \cite{duale2004method} must also be dealt with, which stems from the data dependence and guard conflict between transitions in the path. Accordingly, in order to derive feasible test cases that satisfy the specific coverage criterion, a test generation algorithm must address several key issues as follows: (1) construction of the executable EFSM under test; (2) acquisition of various model information; (3) determination of whether a test path is feasible (executable). 

Constructing the executable EFSM in memory is the basis for test case generation. In this context, the executability of the EFSM refers to the simulation of the dynamic transfer behavior of model transitions and the update actions of the relevant model variables. A common approach is to use manual hard-coding \cite{yang2015efsm} to translate the system specification into an executable EFSM using a specific programming language. But this approach can be time-consuming and error-prone if several different models are tested. SMC \cite{smc} is able to automatically obtain an executable EFSM, but it does not have access to more detailed information about the model. Additionally, test generation algorithms often require a mechanism to flexibly obtain various information about the model such as the use of variables or statistics on variables and transitions. In especially, in data flow-based test generation, it is necessary to rely on these acquired model information to specify test coverage targets (e.g. c-use coverage, p-use coverage, etc.) \cite{el2017assessment}. Although the need for model information extraction can be achieved by static analysis or hard-coding \cite{zhao2020diversity}, this also increases the overhead of the test algorithm implementation and is error-prone. Finally, whenever a candidate test path is generated, it needs to be determined if the path is truly executable. Human-assisted test path 
feasibility discernment significantly reduces the efficiency of test generation. The hard-coded approach  has the limitation of inflexibility. Using JEval \cite{yang2011improve} is a way to work out whether the transition is executable or not, but you still need to get the variable name to assign to the tool. Therefore, to address the above issues, it would be effective way to develop a test generation tool to provide these basic services for the test generation algorithm. By this means, test algorithms can only focus on implementing the business logic for test case generation according to test coverage 
criteria, thus improving the quality and automation of test case generation.

In this paper, we propose a new automated test generation tool, PTSG, to provide underlying services for the implementation of test algorithms, making test case derivation more convenient and efficient. It automatically builds an executable EFSM model in memory based on the system specification and opens service interfaces to interact with the model. By relying on these services provided by the tool, the business logic in the test generation algorithm can be decoupled from the cumbersome interaction with the model. Consequently, a tester can concentrate on implementing the test case generation algorithm itself, without having to consider the details of the operations interacting with the model. More specifically, The initial input to the tool is an EFSM description file described by the tester using JSON syntax. Using this file, PTSG performs lexical and syntactic analysis using a custom parsing engine to construct the various objects and interaction methods associated with the EFSM. This parser-based approach allows us to automatically extract model information (variable names, types, initial values, etc.) at a very small granularity. Moreover, compared to the manual coding approach, the method not only increases the reusability of the model, but also significantly reduces the overhead of algorithm implementation. Finally, by calling these model manipulation interfaces, the test algorithm can easily obtain the desired model information and implement automatic determination of test path executability. PTSG$ \footnote{https://github.com/luckyPer/PTSG} $ is open source and can also be easily extended with 
various test generation algorithms. Without loss of generality, these functions provided by our tools can also serve the implementation of other related methods, such as mutation testing, test case prioritization, etc. 

The rest of this paper is organized in the following sections: Section II briefly presents related work. Section III provides some basic definitions used in our tool. Section IV introduces an overview of the tool and its implementation detail. Then Section V gives a case study to illustrate the usability and effectiveness of the tool. Finally, Section VI concludes this paper and suggests the future work.

\section{Preliminaries}
In this section we present background knowledge on EFSM-based test suite generation to facilitate understanding of our proposed tool. First, we introduce the relevant definitions of EFSM models. Then, the usefulness of an executable EFSM model for testing is briefly described. Finally, a form to follow when parsing model specifications is presented.
\subsection{EFSM model and related definitions}
An extended finite state machine (EFSM) \cite{petrenko2004confirming} can be described as a 6-tuple {\itshape M} = $\langle${\itshape S}, {\itshape $s_{0}$}, {\itshape V}, {\itshape I}, {\itshape O}, {\itshape T}$\rangle$. Where {\itshape S} represents is a finite set of states. {\itshape $s_{0}$} $\in$ {\itshape S}, is the initial state of the model. {\itshape V} is the set of all variables in the model, containing the context and input variables of the model. {\itshape I} is the input event of the model, which can receive input variables, e.g. ``{\itshape {\rm ?}ip.input}({\itshape $v_{1}$}, {\itshape $v_{2}$}, ..., {\itshape $v_{n}$})", representing an input event of the node {\itshape ip}, which receives input variables from {\itshape $v_{1}$} to {\itshape $v_{1}$}. {\itshape O} is the output event of the model, e.g. ``{\itshape {\rm ?}o.output}({\itshape $v_{1}$}, {\itshape $v_{2}$}, ..., {\itshape $v_{n}$})", which represents the output event of the node {\itshape o} and outputs the variables $v_{1}$ to $v_{n}$. {\itshape T} represents the set of variables contained in the model, each of which can also be represented as a six-tuple {\itshape t} = $\langle${\itshape $s_{h}$}, {\itshape i}, {\itshape o}, {\itshape p}, {\itshape a}, {\itshape $s_{t}$}$\rangle$. Where {\itshape $s_{h}$} and {\itshape $s_{t}$} represent the head and tail states of transition {\itshape t}; {\itshape i} $\in$ {\itshape I} and {\itshape o} $\in$ {\itshape O} are the input and output events of the current transition, respectively. {\itshape p} is the predicate of the transition, which is used to determine whether the current transition is executable or not. {\itshape a} represents an update operation of a transition on variables. The test case of EFSM is composed of test path and test data. 

{\bfseries Definition 1} (Adjacent Transition Pair)
The adjacent transition pair is that, there are two adjacent transitions {\itshape $T_{i}$} and {\itshape $T_{j}$}, where the tail state of {\itshape $T_{i}$} is the head state of {\itshape $T_{j}$}. A test path is a sequence of several adjacent transition pair, i.e., {\itshape P} = $\langle${\itshape $T_{1}$}, {\itshape $T_{2}$}, ... ,{\itshape $T_{i}$}, {\itshape $T_{j}$}, ..., {\itshape $T_{n}$}$\rangle$. 

{\bfseries Definition 2} (Feasible path)
A feasible path is one in which there exists at least one set of input variable values such that the predicate in all transitions in path {\itshape P} are satisfied, where the variable values that trigger the execution of this set of paths are called test data. 

{\bfseries Definition 3} (State Configuration). A state configuration represents the current situation of an EFSM model. A state configuration is a combination of the current state and variable values and can be represented as a two-tuple {\itshape $sc_{i}$} = $\langle${\itshape $s_{i}$}, {\itshape $v_{i}$}$\rangle$, where {\itshape $s_{i}$} $\in$ {\itshape S} and {\itshape $v_{i}$} = ({\itshape$v_{1}$}, {\itshape $v_{2}$}, ..., {\itshape $v_{n}$}) is an n-dimensional value vector where the variables {\itshape $v_{1}$}, {\itshape $v_{2}$}, ..., {\itshape $v_{n}$} $\in$ {\itshape V}.
\begin{figure}
\centerline{
\includegraphics[width=1\linewidth]{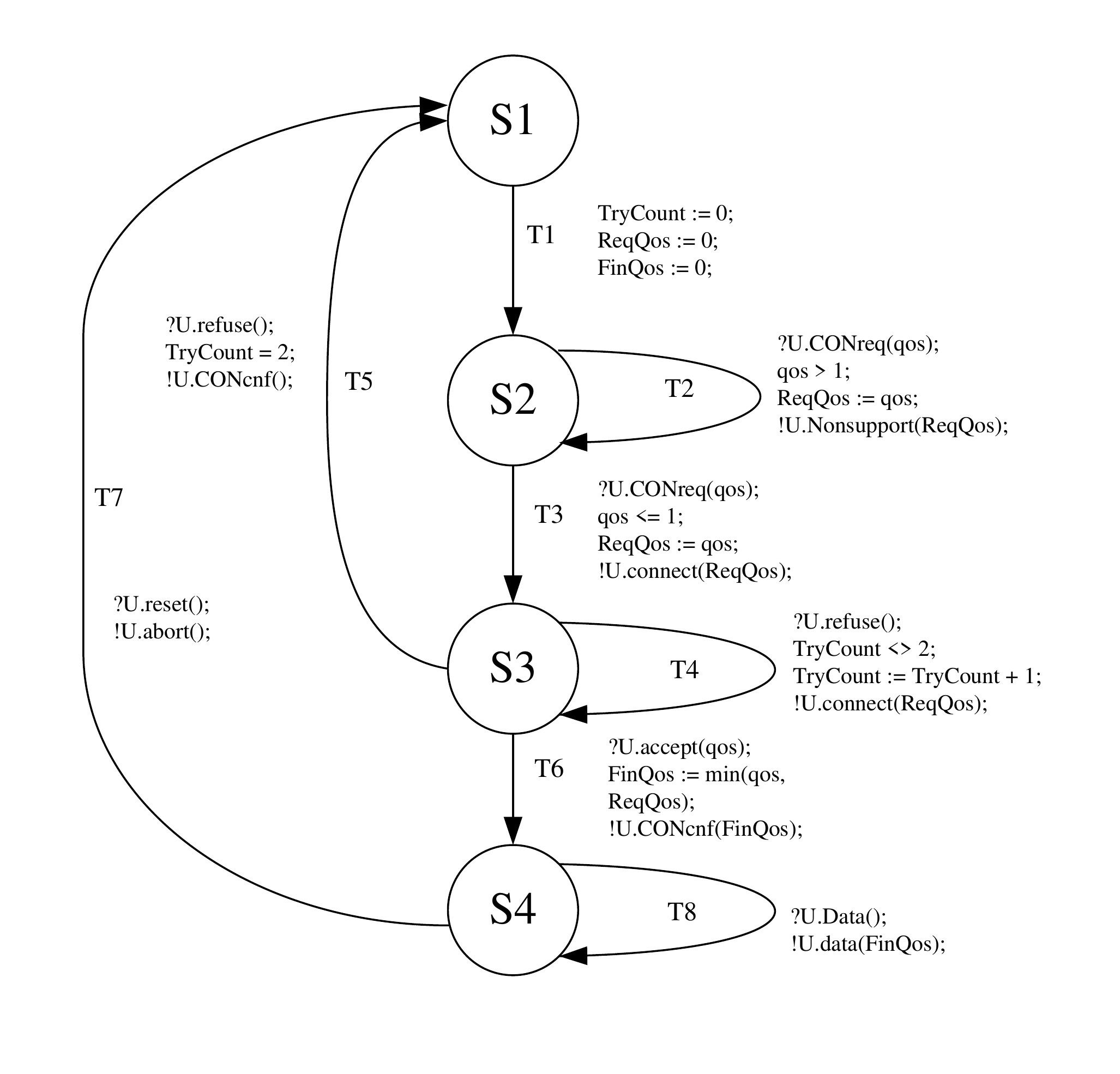}
}
\caption{An EFSM model for Simple Connectivity Protocol}
\end{figure}

Figure 1 shows, as an example, an EFSM model for the Simple Connectivity Protocol (SCP) \cite{cavalli2003new} , which allows us to connect an entity called the upper layer to an entity called the lower layer. The SCP contains eight transitions from {\itshape }$T_{1}$ to {\itshape }$T_{8}$ and four states from {\itshape }$S_{1}$ to {\itshape }$S_{4}$. The set of variables is {\itshape V = {\rm (}TryCount, ReqQos, FinQos, qos{\rm )}}, each of which is of type integer. The first three of {\itshape V} are context variables and the last one is an input variable. For example a path {\itshape P} = $\langle$$T_{1}$, $T_{3}$$\rangle$ has an initial state configuration $\langle${\itshape }$S_{1}$, {\rm (}0, 0, 0, 0{\rm )}$\rangle$ and when passing through {\itshape }$T_{1}$ the state configuration is $\langle${\itshape }$S_{2}$, {\rm (}0, 0, 0, 0{\rm )}$\rangle$.
If the variable value of {\itshape qos} in the input event {\itshape {\rm ?}U.CONreq{\rm (}qos{\rm )}} is 1 at this point, the predicate in {\itshape }$T_{3}$ is satisfied and the tail state configuration becomes $\langle${\itshape }$S_{3}$, {\rm (}0, 1, 0, 1{\rm )}$\rangle$ after the action executed in {\itshape }$T_{3}$.

\subsection{Executable EFSM}
Unlike the general static model, an executable EFSM model \cite{yano2011most} is derived from a static model specification and can be executed to dynamically simulate the model behaviour. This means that the executable EFSM can act like program code to trigger the guards and actions defined in the transition and output runtime information. With the executable EFSM, it helps us to exclude the problem of infeasible test paths during test generation and collect runtime feedback to validate the correctness of the generated paths. It is common to construct dynamic executable model by semantic analysis of expressions in the transition \cite{yang2011improve}. In the test case generation, whenever a transition is traversed, the guards are parsed and combined with the current state configuration to derive a flag for feasibility, and then the actions are executed to obtain the test path, test data and output results. That is to say only those transitions that hold the current guard during the execution of the model are selected as candidate paths. Therefore, the test results can be easily obtained and verified through the executable EFSM.

\subsection{Parsing grammar rules of EFSM}
The construction of an executable EFSM requires syntactic parsing of the static model description. Grammar, as a rule specifying the structure of the language, is content-independent. The Extended BackusNaur Form (EBNF) \cite{laros2011formalized} is used to describe a formal system of syntax and is a meta-language notation for context-free grammar. EBNF decomposes a grammar into a set of rules which describe how the tokens of a programming language form different logical units and is a specification for the definition of a set of computer language symbols in a recursive way. 

The basic structure of the grammar rule is ``$\langle${\itshape non-terminal}$\rangle$::=$\langle${\itshape terminal1}$\rangle$ $|$ $\langle${\itshape terminal2}$\rangle$", where some reserved words and uninterpretable meta-symbols (e.g. numbers, letters, comparison symbols, etc.) are called terminals. $\langle${\itshape non-terminal}$\rangle$ indicates non-termination symbol, meaning that it can continue to be interpreted and replaced by the terminal on the right as required. The symbol ``::=" means ``as defined" and is followed by a series of terminals and non-terminals. The symbol ``$|$" represents that the rule can have multiple interpretation options, i.e. the meaning of or.
\begin{figure}
\centerline{
\includegraphics[width=1\linewidth]{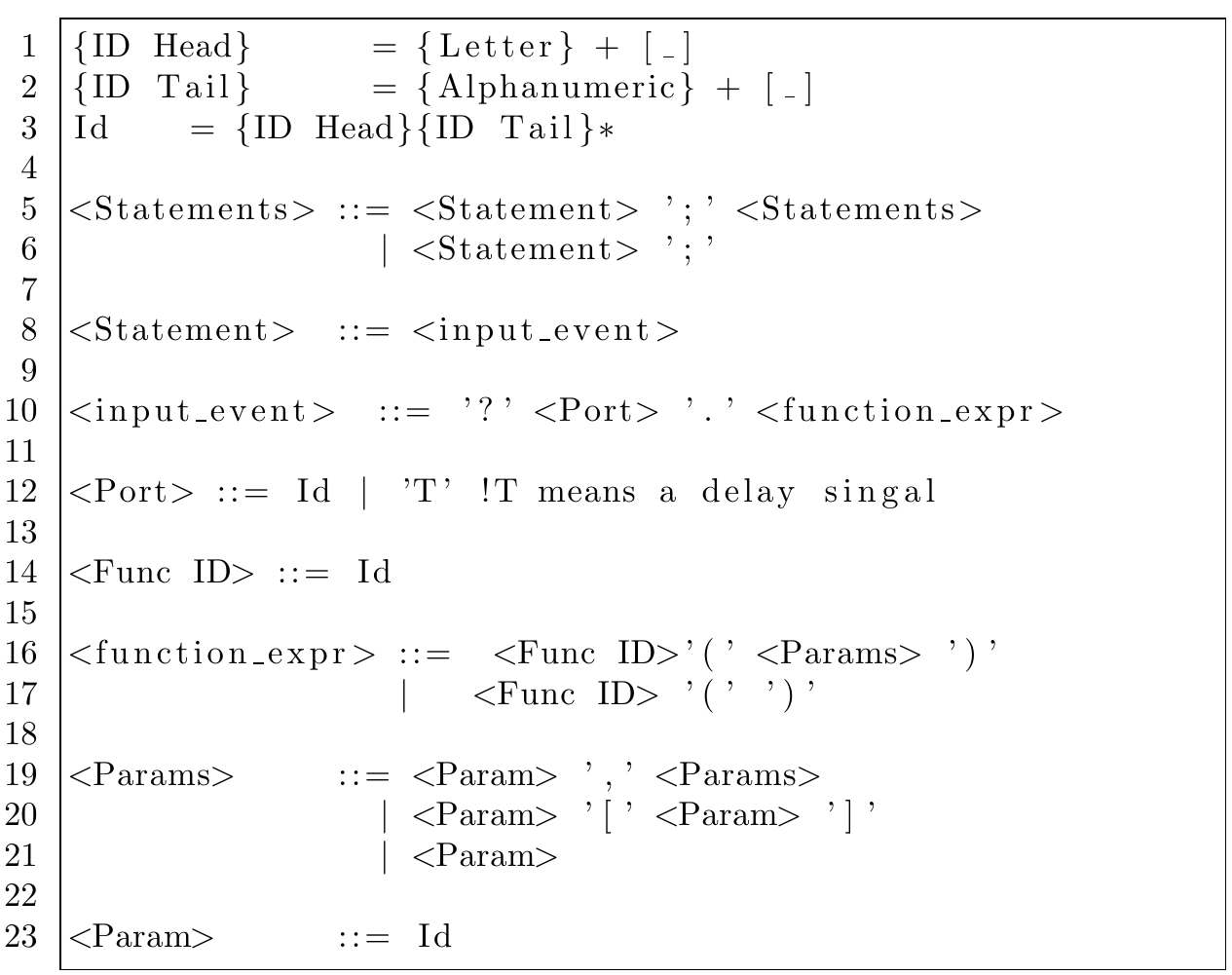}
}
\caption{An example of EBNF grammar}
\end{figure}

Figure 2 shows an example of applying EBNF to input event in the transition of the EFSM model. The $\langle${\itshape input\_event}$\rangle$ rule is a non-terminator, which is a combination of the two termination symbols (consisting of ``?" and ``.") and two non-termination symbols (comprising $\langle${\itshape Port}$\rangle$ and  $\langle${\itshape function\_expr}$\rangle$). The $\langle${\itshape Port}$\rangle$ can be interpreted by a ``T" or the symbol ``Id", which represents some combination of letters and numbers. 

\section{Overall Framework}
In this section, we present the overall framework of PTSG. As shown in Figure 3, the EFSM model specification is first taken as input, the model is parsed by the parsing engine to obtain the parse grammar tree of the model, then the tree is used to construct the EFSM object, and finally the test case is output as the result according to the set generation algorithm and coverage criteria. Figure 4 illustrates the partial class diagram structure of the tool. The EFSM class contains the transition class for edge information, the state configuration class for the current situation and a series of service interfaces (highlighted part of the service interfaces) that can be called. The Parser class parses the model specification and returns the node tree structure to the EFSMParser class. The EFSMParser class is the intermediate interaction class between the EFSM class and the Parser class, through which the node tree is constructed into EFSM object.
\begin{figure*}
\centerline{
\includegraphics[width=1\linewidth]{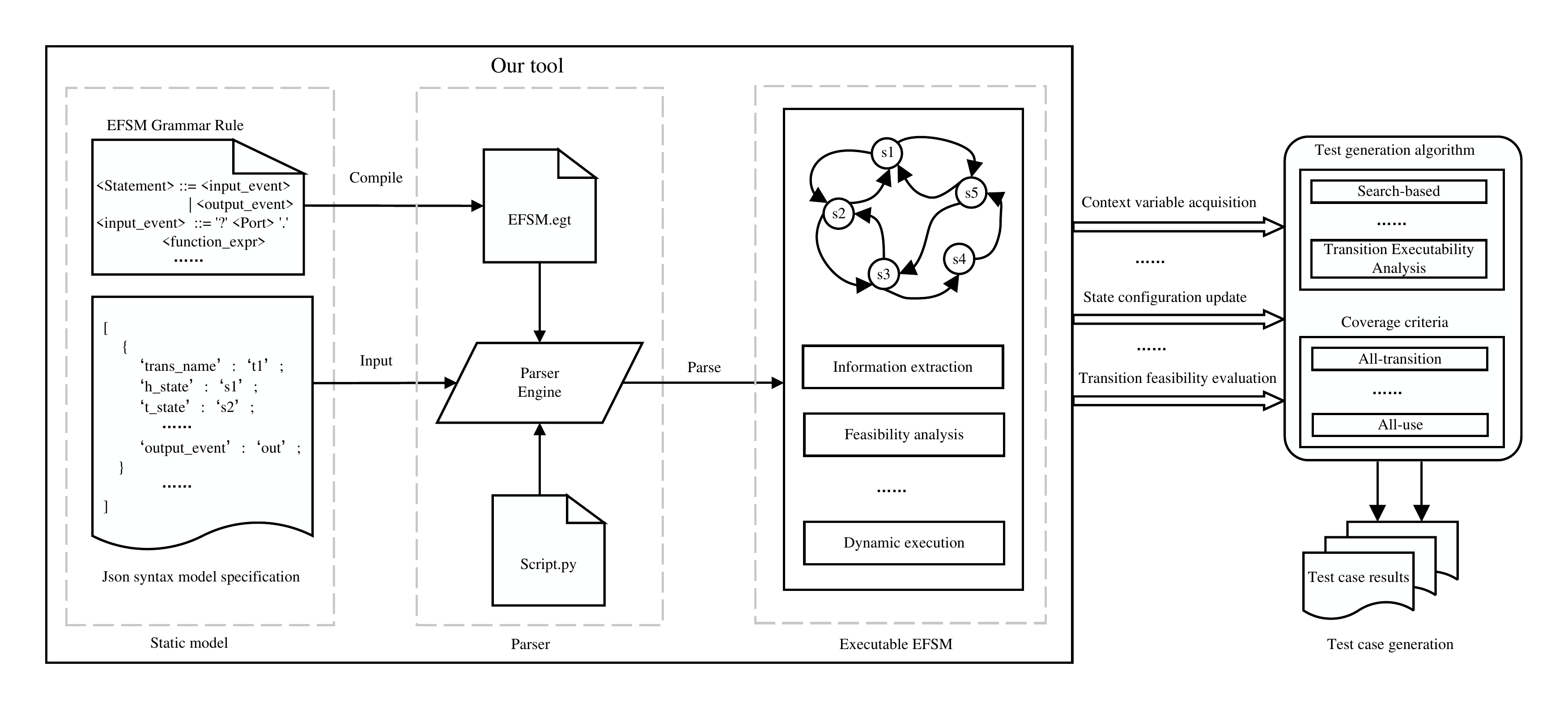}
}
\caption{Overall framework}
\end{figure*}
\begin{figure*}
\centerline{
\includegraphics[width=1\linewidth]{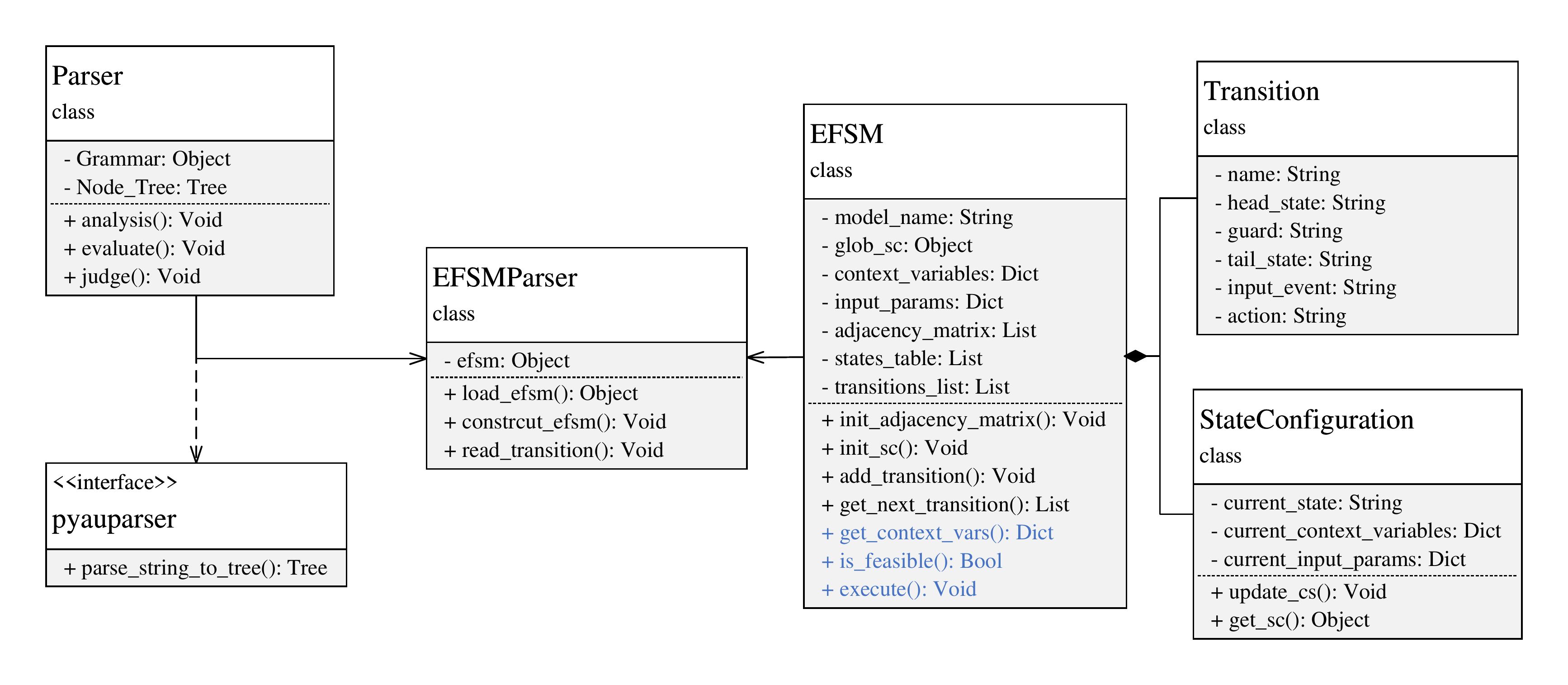}
}
\caption{Partial Main Classes}
\end{figure*}

\subsection{Model Description}
As mentioned above, the model specification will be used as input to the tool. How can the EFSM model be described concisely and clearly in programming language? Here we consider using JSON syntax, which is not only easy to read and write, but also easy for programs to parse. Furthermore, the EFSM model is an abstraction of a real program, and we can describe the whole EFSM model in terms of transitions and state nodes. A transition in the model description file consists of seven keys: transition name, header state, tail state, input event, guard, action and output event. As shown in Figure 5, this is an example of the representation of a single transition in the EFSM model. It represents transition 4 in the SCP model, and the values of the attributes in the format should satisfy the actual meaning of the transition. The whole EFSM model can be represented by an array of such transitions. All the user has to do is to create a new JSON file in the folder consisting of the above and customise the protocol model as required. 
\begin{figure}
\centerline{
\includegraphics[width=1\linewidth]{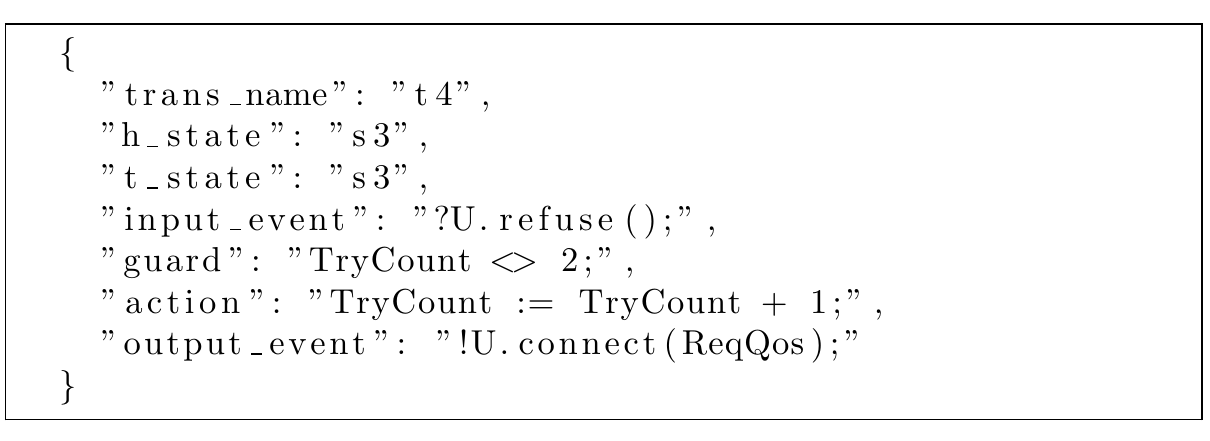}
}
\caption{Model description of a transition in JSON syntax}
\end{figure}

\subsection{Parser}
In EFSM-based test generation, it is necessary to implement feasibility analysis and dynamic execution of the model. This means that we need to extract model information from the previously defined description in JSON syntax and exercise predicate and execute action in transitions. To achieve this, we resort to Gold parser \cite{Goldparserbuilder} and PyAuParser \cite{PyAuParser} to construct executable model. Gold parser is an open source interpreter that compiles grammar rule files following EBNF into .egt files. PyAuParser is able to read .egt files to parse strings or files to node tree.
\begin{figure}
\centerline{
\includegraphics[width=1\linewidth]{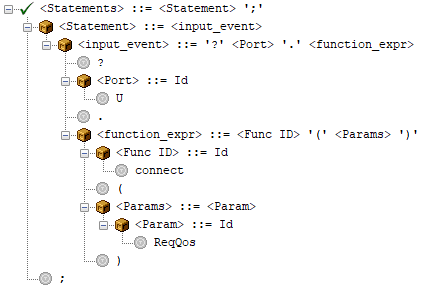}
}
\caption{Parse grammar tree (Screenshot)}
\end{figure}
Figure 2 shows the grammar file for parsing input events in EFSM, specifying that the input event starts with symbol ``?" followed by a function expression under the node, the function contains a number of arguments, which are parsed recursively when there are multiple arguments, and the input event ends with symbol ``;". When an input event statement is received, the parsing engine will read it from left to right and progressively obtain the token. As an example, Figure 6 shows the parse grammar tree generated by using the parsing engine for the input event ``{\itshape {\rm ?}U.connect{\rm (}ReqQos{\rm )};}", representing the function connect under the node {\itshape U}, containing the parameter {\itshape ReqQos}. In the parse grammar tree, {\itshape $\langle$Statements$\rangle$} represents the start symbol, which is further interpreted by {\itshape $\langle$Statement$\rangle$}, and the {\itshape $\langle$input\_event$\rangle$} token is the statement we want to start parsing, which consists of the terminator ``?" and ``." , the non-terminator {\itshape $\langle$Port$\rangle$} and {\itshape $\langle$function\_expr$\rangle$}. The {\itshape$\langle$function\_expr$\rangle$} needs to be further interpreted as {\itshape$\langle$Func ID$\rangle$ `{\rm (}' $\langle$Params$\rangle$ `{\rm )}'}, where the {\itshape$\langle$Params$\rangle$} is recursively reduced to the terminator.
The parsing steps for the model are summarised as follows: (1) use the gold parser interpreter to compile the grammar files based on EBNF to produce .egt files.
(2) read in the statements to be parsed and use the parsing engine to transform them into a parse grammar tree.

\subsection{Executable model construction}
Model construction involves converting the model description into a dynamic EFSM object. The content of each transitions is read from the JSON syntax and further instantiated into a specific transition object. An adjacency matrix is then constructed by adding the transition object to it at the corresponding position using the head state and tail state of the transition. The extraction of context and input variables from the EFSM model is also an essential step for the executable model. The table of variables is constructed by obtaining the values of action, input event and guard from the transitions and initialising the variables according to their types. The default state is set to {\itshape $s_{1}$}, which is constructed as a state configuration along with the previously parsed variable table, making it part of the EFSM object to indicate the current situation. In addition to this, the most critical step is the feasibility analysis of the guards and the execution of the actions in the transitions. Our tool first obtains the current guard in the transition and convert it into a parse tree. The parse tree is then traversed and when variables are encountered, the corresponding values are found in the context variable table or input variable table, and the values of the variables in the statement are compared to return a boolean value of whether the statement is feasible or not. If the current transition is executable, the new value of the context variable is written back to the context variable table through the same parsing and traversal steps, then moved to the next state and finally the state configuration is updated.

\subsection{Test generation}
With the EFSM object constructed as described above, test cases can be generated according to the test generation algorithm after the test coverage criteria has been determined. EFSM based test case generation consists of test data and test path generation. The test path generation is to find the set of paths that can satisfy the selected coverage criteria, while the test data is to trigger the paths (i.e. the current given transition) during the execution of the model, mainly to find the input variable values of the input events in the transition. For example, search-based genetic algorithms \cite{kalaji2011integrated} focus on the computation of the fintness function, and Transition Executability Analysis \cite{shu2016heuristic} focuses on how to mitigate the state explosion, but these algorithms cannot be separated from the executable EFSM. Our tool can provide functions such as transition feasibility analysis and model information extraction to different generation algorithms in the form of interfaces, in order to avoid the underlying information processing of the model and to focus solely on the specific generation algorithm logic. Furthermore, users can extend the functionality based on the information already extracted from variables, transitions, inputs etc. under the EFSM class to realise data flow based coverage criteria.

\section{Case Study}
\begin{table*}
	\centerline { Table 1. Generation Results of SCP model }
        \vskip2pt
        \tabcolsep=0.5cm
	\renewcommand\arraystretch{1.5}
	\begin{tabular}{|l|l|c|l|}
		\hline
  
		\thead{\bf Test Case No} & \thead{\bf Test Path} & \thead{\bf The tail State Configuration \\ \bf (State, TryCount, ReqQos, FinQos, qos)} & \thead{\bf Output} \\ \hline
  
		\multirow{3}{*}{TC 1}
            & `$t_{1}$' & (`$s_{2}$', 0, 0, 0, 0) & -  \\ 
		& `$t_{1}$', `$t_{2}$' & (`$s_{2}$', 0, 2, 0, 2) & !U.Nonsupport(2); \\  
            & `$t_{1}$', `$t_{2}$', `$t_{3}$' & (`$s_{3}$', 0, 1, 0, 1) & !U.connect(1); \\ 
            \hline

		\multirow{3}{*}{TC 2}& `$t_{1}$' & (`$s_{2}$', 0, 0, 0, 0) & - \\ 
		& `$t_{1}$', `$t_{2}$' & (`$s_{2}$', 0, 2, 0, 2) & !U.Nonsupport(2); \\  
            & `$t_{1}$', `$t_{2}$', `$t_{2}$' & (`$s_{2}$', 0, 2, 0, 2) & !U.Nonsupport(2); \\ 
            \hline

  		\multirow{3}{*}{TC 3}& `$t_{1}$' & (`$s_{2}$', 0, 0, 0, 0) & -  \\ 
		& `$t_{1}$', `$t_{3}$' & (`$s_{3}$', 0, 0, 0, 0) & !U.connect(0); \\  
            & `$t_{1}$', `$t_{3}$', `$t_{6}$' & (`$s_{4}$', 0, 0, 0, 0) & !U.CONcnf(0); \\ 
            \hline   
	\end{tabular}
\end{table*}
In this section, we use PTSG to generate test case that satisfy a specific coverage criteria and perform a case demonstration of the generated content. We use the typical Simple connection protocol \cite{cavalli2003new} (SCP) model as the subject of our experiments.  The detail of test generation is shown in Pseudocode 1: 

\begin{algorithm} 
  \SetAlgoLined
  \renewcommand{\algorithmcfname}{Pseudocode}
  \KwIn{EFSM Specification M}
  \KwOut{Test Suite T}
    EFSM = EFSMParser(M)\;
    C = set\_coverage\_criteria()\;
    ${SC}_{0}$ = EFSM.get\_init\_sc()\;
    put ${SC}_{0}$ into T\;
    \Repeat{T Satisfying C}{
        \While{T is not empty}{
            previous\_sc = T.get\_sc()\;
            T.pop()\;
            current\_state = previous\_sc.get\_cur\_state()\;
            current\_transition = EFSM.get\_next\_tran(
            
            \qquad current\_state)\;
            input\_data = generate\_data()\;
            context\_vars = previous\_sc.get\_cur\_context()\;
            \If{EFSM.is\_feasible(current\_transition)}{
                state = goto\_next\_state()\;
                new\_sc = EFSM.update\_sc(input\_data,

                \qquad context\_vars, state, current\_transition)\;
                put new\_sc into T\;
            }
        }
    }
    \caption{Test Generation}
\end{algorithm}

\begin{enumerate} 
 \setlength{\itemsep}{-2ex}  
 \setlength{\parskip}{0ex} 
 \setlength{\parsep}{0ex}
\item Input the model specification into the class EFSMParser to obtain the EFSM object;\hfil\break
\item Set a coverage criteria and traverse the adjacency matrix from the state in the obtained initial state configuration to continuously expand the TEA tree;\hfil\break
\item For each transition, the current input variables are obtained and passed into the interface provided by the EFSM object to check if current transition is executable to expand the tree; \hfil\break
\item If current transition is executable, continue to call the operation to update the state configuration;\hfil\break
\item Discard the current test path if it is infeasible;\hfil\break
\item Determine if the generation is finished based on the selected test coverage criteria.
\end{enumerate}

The test path algorithm used in this section is Bread-First-Search (BFS) \cite{kozen1992depth}, the test data generation algorithm is boundary value analysis \cite{ramachandran2003testing}, and the test coverage criterion is all state coverage \cite{devroey2014coverage}. Table 1 shows a set of test suite that satisfy the all state test coverage criteria, which is to cover all states in the SCP model. The initial state configuration is (`{\itshape }$s_{1}$', 0, 0, 0, 0). In test case 1, after transition {\itshape }$t_{1}$, state {\itshape }$s_{2}$ is reached, at which point the state configuration changes to (`{\itshape }$s_{2}$', 0, 0, 0, 0, 0). When the value of the input variable {\itshape qos} is received as 2, the gurad of the transition {\itshape }$t_{2}$ is satisfied, the context variable {\itshape ReqQos} is updated to 2, and the output is ``!{\itshape U.Nonsupport}(2);", so the state configuration at this point is (`{\itshape }$s_{2}$', 0, 2, 0, 2). The input variable {\itshape qos} is then received with a value of 1, so that the transition {\itshape }$t_{3}$ is executable, and as the context variable {\itshape ReqQos} is updated to 1, the output at this point is ``!{\itshape U.connect}(1);". The state covered by test case 1 at this moment is ({\itshape }$s_{1}$, {\itshape }$s_{2}$, {\itshape }$s_{3}$). With the generation of test case 3, state 4 is covered, thus satisfying the all state coverage requirement, and the generated test suite is (TC1, TC2, TC3).

\section{Related Work}
SMC \cite{smc} is a state machine compiler application, designed primarily in Java. This tool compiles state machine description in sm syntax into various programming language code. The generated state machine class and the transition logic, together with user's own business code logic, can be used to implement an executable state machine. However, the code generated by SMC can only acquire a small amount of information about state, changes, etc. and cannot satisfy some data flow-based dependencies (e.g. def-use chains, P-use coverage).

Yang et al. \cite{yang2011improve} aims to improve the effectiveness of test case generation on EFSM through path feasibility analysis. They use JEval to parse and evaluate the expressions of the EFSM model and make the EFSM executable. Unfortunately, since the code is not open source, we do not have access to the data structure and implementation of the model in the article.

TCG \cite{muniz2015model} is a model-based tool for the generation and selection of functional and statistical test cases. It was developed as a plug-in to LoTuS and as TCG handles formalisms based on state machines (LTS and PLTS).And it can only provide test case generation that satisfies the transition based test coverage criteria.

TAF \cite{robert2021taf} is an automated test case generation tool based on the xml data model. It generates various test cases by combining random sampling and the use of SMT solvers. An xml file is written to describe the model under test using the node attribute of the tag pair. The model is then represented by parsing the file to create a tree structure, which also increases the complexity of writing the model description.

JSXM \cite{dranidis2012jsxm} is an automated test case generation tool that follows a kind of EFSM (Stream X-machines) and is developed in Java. The model specification is described in XML-based language and the function is written in Java inline code. It is relatively cumbersome to construct the model, and some details of the model cannot be obtained in the test generation.

Chimisliu et al. \cite{chimisliu2012model} propose a method for automatically converting systems consisting of communication state diagrams into a language of temporal specifications (LOTOS). They show how test cases can be generated in a semi-automatic manner by using user input as interpretation of UML diagrams. The generation process is less effective when the annotate states and/or transitions provided partially describing a test scenario are rejected.

In contrast, our tool is capable of constructing executable models to simulate the dynamic behavior of the system. PTSG can also extract the details of the model through parsing to meet different coverage criteria.

\section{Conclusions And Future Work}
In this paper, we introduce an open source EFSM-based test generation tool, PTSG. Our tool constructs a dynamic executable EFSM model by using a custom parsing engine to perform lexical and syntactic analysis of the model specification in JSON syntax. Based on this approach, we can dynamically implement simulations of model behaviour and automatically obtain information about the model in a fine granularity. In addition, our tool provides a rang of interfaces to call such as information acquisition, transition feasibility assessment and model simulation execution. In this way, the tester can be separated from the tedious model interaction and the implementation of the test algorithm. We have chosen a model specification for the case study to automatically generate test cases that meet selected test coverage criteria and the results show that the content generated by our tool is convenient and valid.

Our research is still preliminary and there is much work to be done in the future. We hope our tool to incorporate mutation testing, to evaluate the generated test suites by mutating the model specification, and to turn PTSG into a more integrated testing tool.


\section*{Acknowledgment}
This work was supported by the Zhejiang Provincial Natural Science Foundation of China under Grant No.LY22F020019, the Zhejiang Science and Technology Plan Project under Grant No.2022C01045, and the Natural Science Foundation of China under Grants 62132014 and 61101111. 




%

\smallskip

\end{document}